# A Colored Petri Net Model of Simulation for Performance Evaluation for IEEE 802.22 based Network

Eduardo M. Vasconcelos[1] and Kelvin L. Dias[2]
[1]*Federal Institute of Education, Science and Technology of Pernambuco, Garanhuns, Department of Informatics*
*e-mail address: Eduardo.vasconcelos@garanhuns.ifpe.edu.br*
[2]*Federal University of Pernambuco, Center of Informatics*

Cognitive Radio is a new concept that allows radio devices access to licensed bands since they do not cause harmful interferences to systems that hold the license of use. The main motivation for the increase of research on Cognitive Radio is the scarcity of non-licensed bands due to the large employment of wireless networks on cities. In this paper, we describe a Cognitive Radio Model of Simulation designed through the Colored Petri Net Formalism. This represents an effort to deliver to scientific community a model of simulation that is easily extensible and graphically validated. Through comparison with literature, we have demonstrated that this model is not invalid.

## I. INTRODUCTION

Cognitive Radio (CR) is a concept that allows smart radio devices to access licensed bands opportunistically, among other features [1]. It has become a relevant paradigm due to the scarcity of non-licensed bands and the popularization of wireless networks. However, the main conern regarding opportunistic access is the protection of the legacy system's communication, also known as Primary Users (PUs).

As long as the interference to the PU is kept beyond a defined threshold, the opportunistic systems that are commonly referred as Secondary Users (SUs) may operate in licensed bands. This implies that the SUs should detect PU transmissions in order to avoid interference [2]. One way to enhance PU detection is to learn its behavior through spectrum sensing techniques [2].

The IEEE 802.22 standard [3] has introduced the cognitive concept for Regional Area Network environments, leading Internet access to the last mile users. In this standard, one backup channel is previously selected in order to reestablish the secondary communication after a PU is detected.

The problem that we are addressing in this paper is the scarcity of simulators that represents the events of opportunistic access, given that the majority of the works do not properly present their simulators.

In this paper, we have designed a CR model of simulation.

It is based on the Colored Petri Net (CPN) Formalism, using a free application called CPN Tools [4]. The model described in the following chapter represents the set of states and events that represents the interation between SUs and PUs. The advantage of using CPN is the possibility to include new functionalities beyond those developed in this work.

## II. COGNITIVE RADIO MODEL OF SIMULATION

Figure 1 presents the IEEE 802.22 network simulator that was designed using the CPN Formalism. Firstly, three main token objects were developed: *CRNode, CHANNEL and PUInf* that represent the SU, communication channel and PU, respectively. The *CRNode* is composed by four integer values: the SU´s ID, the channel´s ID to which the SU is associated, the SU´s application QoS and the remaining battery charger. Furthermore, the object named *CHANNEL* is characterized by three attributes: the channel id, a Boolean value that represents the presence of a PU and the SU's id that is currently transmitting. Lastly, *PUInf* has four attributes: the channel id to which it is associated, its transmission period, idle period and a boolean value that represents its current state (busy or idle).

The simulation is composed by both transient and steady evaluations. On the transient part, the channels are defined and the SUs arrive in the network. The transient part of the simulation is presented on Figure 2. The transition *Connecting* initializes the simulation by allowing the CPN Tools connect to an external application through the *acceptConnection()* method [4]. We consider the external application as an interesting approach to build a log file of simulation events-Although in this paper we do not specify one code to the external application, we have pre-defined a set of event labels that are sent from the CPN Tools, which can be easily re-defined by other researchers.

The transition *Connecting* is only fired when an external application connects with CPN Tools through port 9000 (used on CPN Tools Examples). When *such transition* is fired, one token with value ‘1’ with 100 time units associated is generated to place “*New CR*”; this value represents the first SU id that is inserted on the simulation. Note that the transition *Connecting* is associated (through an arc)–with the place “*new Channel*”. Also, differently from the arc linking transition *Connecting* and the place “*new CR*”, the existing label on this arc represents a method. That is because in IEEE 802.22 network, the Base Station must select one channel to work as main channel, and this selection is made by the method *SelPrimaryChannel()*. In this simulation, we consider that the bandwidth is divided by the channel ID, e.g. channel 1 represents the bandwidth between 54~60Mhz. So, *SelPrimaryChannel()* returns the channel id of the primary channel.

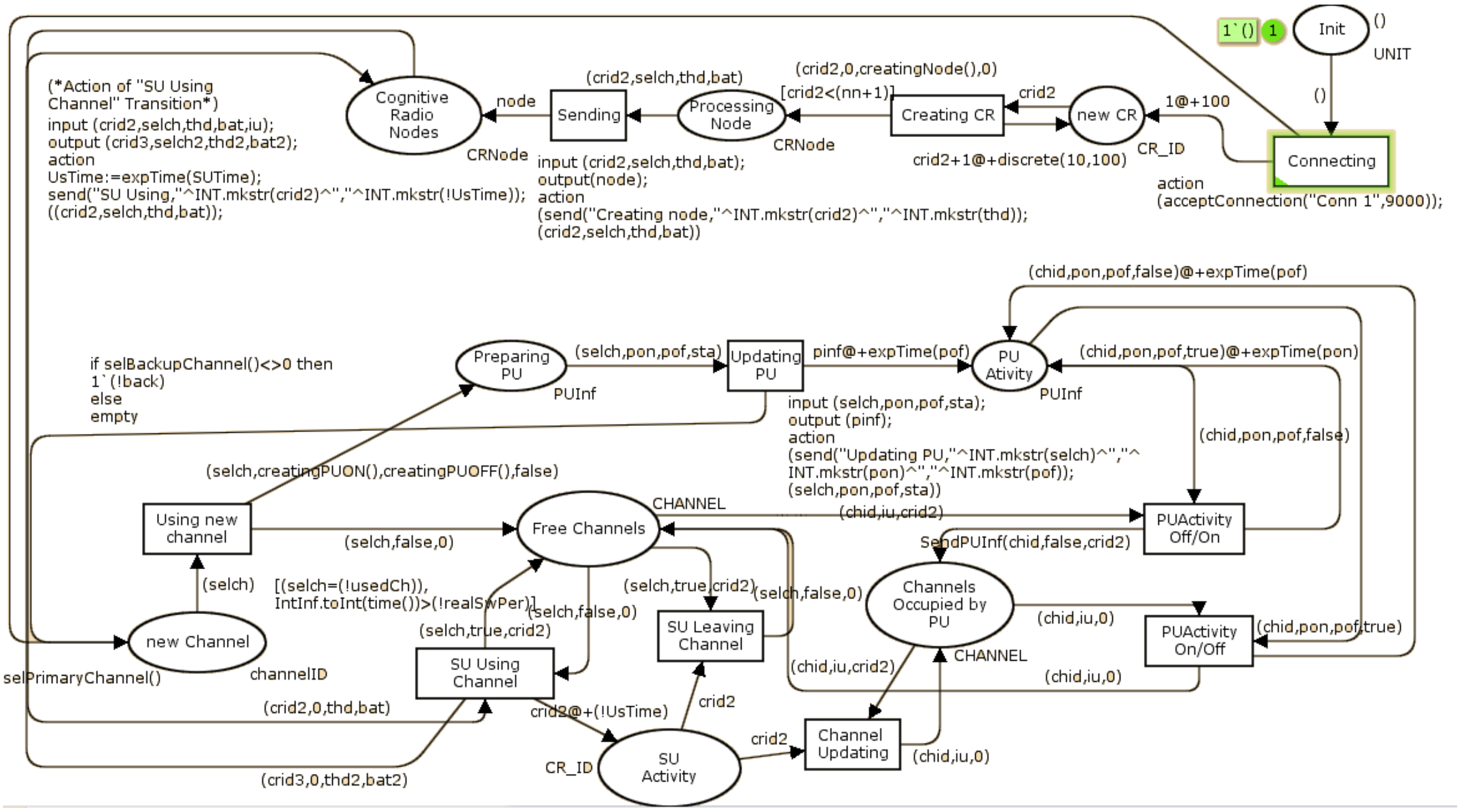

*Figure 1 – Complete Model.*

The process implemented by method *SelPrimaryChannel()* is defined as follows: (1) all channel parameters (ON and OFF periods) are randomly generated; (2) one channel is chosen through a preferred metric, e.g. major channel availability or major off period; (3) the values ON and OFF periods and the primary channel id are stored on the CPN tools variables to be used later by both methods *creatingPUOn*() and *creatingPUOff*(); finally, the method returns to the place “*new Channel*” the selected channel’s id.

After the transition *Connecting* is fired, the place “*new CR*” will have one token containing the value 1 representing first SU id and the place “*new Channel*” will have a token containing the selected id of the main channel. As the token in place “*new CR*” has a *time* associated, the transition “*Creating CR*” will only be enabled when the simulation time reaches the value in that token. In other words, the SUs will only be introduced on the network after the channels are defined. So, firing of the transition “*Using new Channel*” will remove the existing token on place “*new Channel*” generating another token of type *CHANNEL* to the place “*Free Channels*” and one token of type *PUInf* to “*Preparing PU*” as can be seen on Figure 2. Note that all transition firing will not consume simulation time, meaning that this process does not affect any evaluation that is being performed.

To this point, the only enabled transition is “*Updating PU*” that is responsible for creating a PU to the place “PU Activity”. This token contains a period that is randomly defined by an exponential distribution whose average is defined by the OFF period, which means that the PU begins in an idle state.

The transition “*Updating PU*” generates one token to the place “*new Channel*”, where the arc value is defined by a method if the value returned from *selBackupChannel()* is different from 0. In this simulation, we also assumed that the BS considers only two channels, although this characteristic is flexible and can be changed in further works.

The implementation of the method *selBackupChannel()* is similar to *selPrimaryChannel()* method and is described as follows: (1) it is necessary to verify that the local value that represents the backup channel id is defined; if it is true, then the method must return the value 0; (2) the method must verify the list of channels that were previously defined on the method *selPrimaryChannel()* and choose one channel based on the preferred metric, which will be necessarily different from that chosen as the primary channel; (3) it is mandatory to store the ON and OFF periods as well as the backup channel id. If it is desired to use more than two channels on the model, the methods *selPrimaryChannel()* and *selBackupChannel()* should be restructured into a single method that randomly defines each channel. The process of defining channels to the simulation also can be performed by an external application. In order to do so, it is possible to use the CPN Tools methods: *ConnManagementLayer.receive* and *ConnManagementLayer.send*.

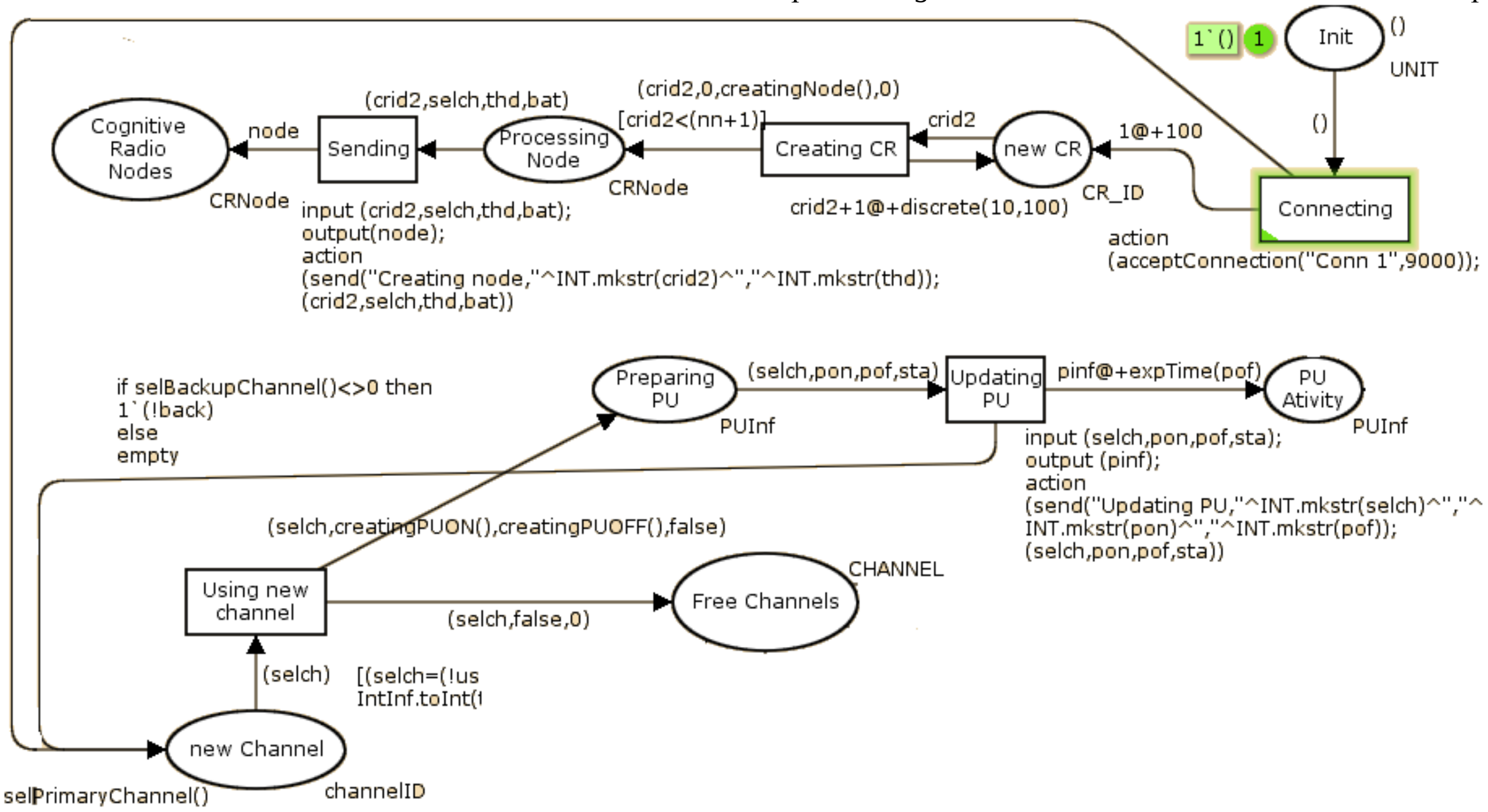

*Figure 2 – Transient part of the Simulator*

Once the channels have been properly defined, the next step is to create the SUs. The transition "*Creating CR*" removes one token *CR_ID* from the place "*new CR*" generating another token of type *CRNode* to the place "*Processing Node*" and other to place "*new CR*". The token generated to the place "*new CR*" represents the id of the next SU that will be created and have a time value randomly associated. The token generated to the place "*Processing Node*" contains the removed id from the place "*New CR*". The method *creatingNode()* is responsible for randomly defining a value for the required data throughput. Although the battery attribute has been defined with 0, any researcher can redefine an initial value. Note that the transition "*Creating CR*" has a guard that disables it when the id of the next SU is greater than the value, which represents the max number of SUs on simulation.

The transition *Sending*, is responsible for sending one message informing the creation of a new SU to the external application. In this transition, we use the *send()* method that is implemented through the API *ConnManagementLayer*. The transient part of simulation ends when the number of inserted SUs is greater than the number defined by the value nn.

The places and transitions of the steady part of the simulation is described in figure 3. The SU's access is defined by transitions "*SU Using Channel*" and "*SU Leaving Channel*". The transition "*SU Using Channel*" makes the association between one SU and one channel. When fired, this transition randomly removes one SU from place "*Cognitive Radio Nodes*" and one channel from place "*Free Channels*". Note that the removed token from place "*Free Channels*" has some pre-defined attributes and the transition "*SU Using Channel*" only becomes enabled if there are tokens with values (_,false,0), where the symbol "_" represents any value, false represents the absence of PUs and 0 represents no SU currently transmitting. Also, there are two guards on the transition "*SU Using Channel*" that becomes enabled if there is a token on place "*Free Channels*" whose channel id is equal to the variable *chUsed* and the simulation time is greater than the value of variable *RealSwPer*. The value of *chUsed* must be defined on the method *selPrimaryChannel()* with the primary channel id and be re-defined when a channel handover is performed. As in this paper we only consider two channels, the *chUsed* value will only assume the ids of the primary and backup channels, but to use more channels on the simulation, it will not be necessary to do any modification to this Variable. The *realSwPer* variable contains the channel switching period and is defined by the following code: *realSwPer*:= *!swTime+IntInf.toInt(time())*, where *swTime* is defined on method *selBackupChannel()* as *swTime := (!CHDiff)* CHBandWidth* swDelPerMHz*, *CHDiff* is the difference among the main and backup channel ids, *CHBandWidth* is the channel bandwidth and *swDelPerMHz* the switching delay per MHz. The code *IntInf.toInt(time())* is used on CPN Tools to return the current simulation time.

The PU occurance on a specified channel is modeled by the transitions "*PU Activity Off/On*" and "*PU Activity On/Off*". Once a PU is created, it has an absence time associated and when this time expires, the two transitions that represent the PU activity may be enabled. The transition "*PU Activity Off/On*" only becomes enabled if the boolean value of PU is false, which represents a PU in the idle state. Otherwise, when the transition is fired, it removes the channel associated with the PU from place "*Free Channels*" and the PU from channel "*PU Activity*", generating one token to place "*Channels Occupied by PUs*"

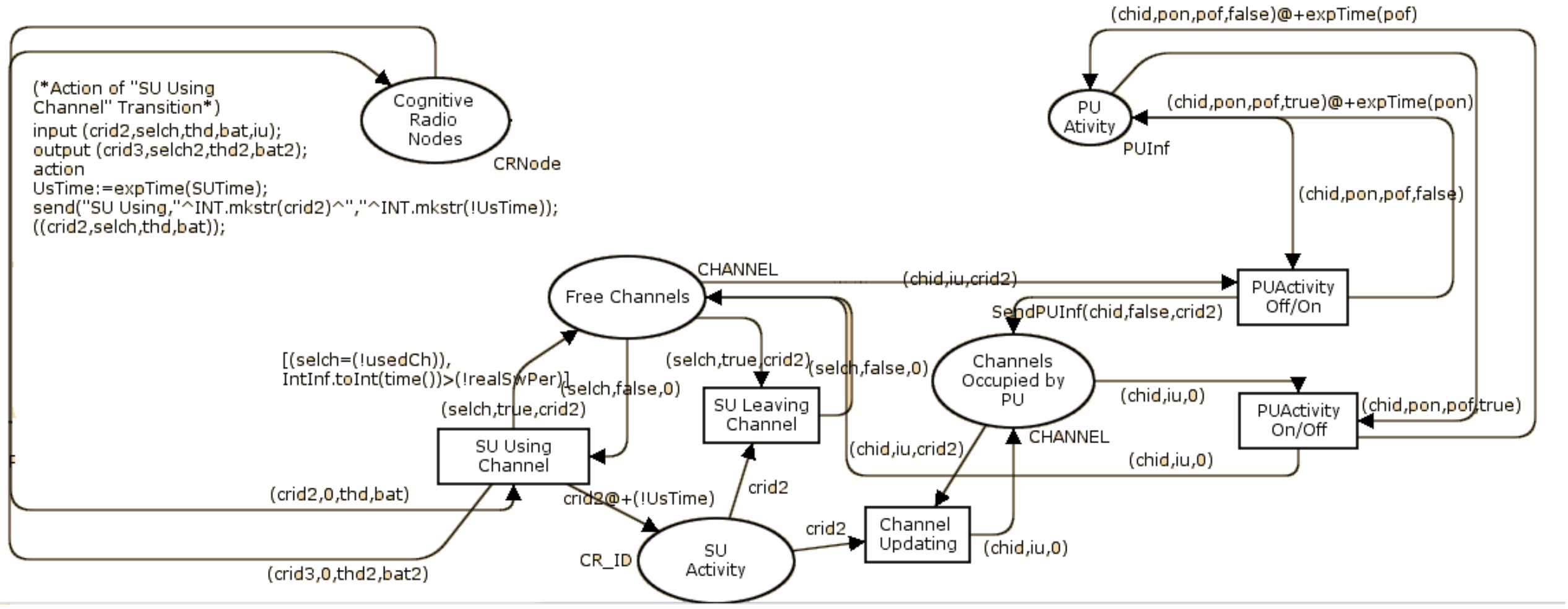

*Figure 3 – Steady Part of Simulator.*

After removing the tokens of places "*Free Channels*" and "*Cognitive Radio Nodes*", the transition "*SU Using Channel*" generates one SU to place "*Cognitive Radio Nodes*", one token to place "*Free Channels*" and one with the SU id that has been assigned to the channel to place "*SU Activity*". This token has a time associated that represents the SU transmission period. The global variable *UsTime* has its value defined on the action declaration of transition "*SU Using Channel*" (top-left corner of Figure 3). This value is defined through an exponential random generator with average value subframe (configured with the subframe duration). In addition, on the action declaration, one message is sent to the external application containing the id of the SU and the transmission duration. To make the SUs transmit for a fixed period, it is necessary to rewrite the label of the arc that connects the transition "*SU Using Channel*" with the place "*SU Activity*", replacing the global variable *UsTime* for a desired value.

Once the time associated with the token of the place "*SU Activity*" expires, the transition "*SU Leaving Channel*" is enabled. By firing this transition, the system removes the SU id from place "*SU activity*" and the corresponding channel from place "*Free Channels*", generating one token to place "*Free channels*" with the channel ready to be used by another SU.

and one token to place "*PU Activity*". The token generated to place "*PU Activity*" has the boolean value modified to true, representing the PU transmission and one time associated that represents the PU transmission period. The ack that connects the transition "*PU Activity Off/On*" calls the method *SendPUInf()*; this method has the task of sending this event to the external application and to change the channel that the Secondary Network is currently operating. This is presented on Algorithm 1.

**Algorithm 1 – SendPUInf() Method on CPN Tools Notation**

```
  fun SendPUInf(cha:channelID,isu:isUsed,cri:CR_ID) =
  let
  in

  send("PU  Off  to  On,"^INT.mkstr(cri)^","^INT.mkstr
(cha));
  if (!usedCh)=cha then
  (
  send("Switching,"^INT.mkstr(!swTime));
  realSwPer:= !swTime+IntInf.toInt(time());
  if cha=(!prim) then
    (usedCh:=(!back);
       (cha,isu,cri)
    )
    else
    (
      usedCh:=(!prim);
      (cha,isu,cri)
    )
```

```
  )
  else
  (cha,isu,cri)
end
```

The method presented on Algorithm 1, follows the CPN Tools formalism. The basic operation of this method is to represent the process of channel switching by change the value of variable *usedCh* to the next channel. Also, this method defines the switching delay by redefining the variable *realSwPer*. If is desired to use more than two channels, it is just necessary to set the value of the variable *usedCh* with one of the channels contained on channel list.

The transition "*Channel Updating*" is used to remove surplus tokens on place "*SU Activity*". It avoids the accumulation of tokens on the place "*SU Activity*" since if the used channel is removed from place "*Free Channels*", the token of the SU that is using this channel will not be removed.

Finally, the transition "*PU Activity Off/On*" represents the same event of transition "*PU Activity On/Off*". Here, differently form the transition "*PU Activity Off/On*", the Boolean value of *PUInf* is changed to true, and the time associated with the generated token is generated based on the PU off period.

## III. MODEL VALIDATION

To validate the proposed simulation model, we have compared results with those presented in [5]. The aim of this evaluation is demonstrate that the model produces comparable results with those obtained by Bayhan and Alagoz [5]. The aim of this validation is not demonstrating that proposed model is valid, but demonstrating that the simulator is not invalid, that is, it represents the $P\{\exists x \in S | f(x) = g(x)\} > 0$, where $S$ is the set of system configuration, $f(x)$ the outcomes from simulation and $g(x)$ the outcomes from the real system.

To perform the simulations, we considered that the Base Station selects the main and backup channels based on the greatest availability calculated by $A = P_{off}/(P_{on} + P_{off})$ with *A* representing the Channel availability. We have used this metric because of the IEEE 802.22 definition [3].

Table 1 presents the parameters of simulation.

**Table 1 – Simulation Parameters [5]**

| Parameter | Value |
|---|---|
| Channel Bandwidth | 5MHz |
| Transmission Power | 1980mW |
| Idle Power | 990mW |
| Circuit Power | 210mW |
| Channel Switching Power | 1000mW |
| Channel Switching Delay | 0.1ms/Mhz |
| Number of Channels | 50 |
| Simulation Period | 3600s |
| Frame Duration | 0.1s |

The metric used is the same used in [5] "*Total Energy Consumption per SU per frame*" in mili joules. On the simulation, this metric has been obtained by equation (1).

$$EnergyConsumedPerJoule = (TransPeriod * TransPower + IdlePeriod * IdlePower + SwitchingPeriod * SwitchingPeriod) * CircuitPower * 0.1/SimTime \quad (1)$$

The time periods are defined in seconds, power variables in mW and SimTime is the simulator period in seconds. We have performed 40 simulation rounds. At the end of simulations the average value of energy consumed per frame has been near of 140mj with an error lesser than 5% considering 95% of confidence. The result is comparable with that achieved by Bayhan and Alagoz [5] that has been approximately 145mj per frame. So, based on results obtained we consider that the model is not invalid.

## IV. CONCLUSIONS

In this paper we have designed a model for the simulation of a Cognitive Radio Network using Colored Petri Net Formalism. After comparing the results with one paper of the literature, we have demonstrated that our simulation model is not invalid.